\documentclass[prc,unsortedaddress,superscriptaddress,onecolumn,amsmath,amsfonts,amssymb,showpacs,floatfix]{revtex4}
\usepackage[dvips]{graphicx}
\usepackage{epsfig}
\begin{document}
\title{Effect of Coulomb Forces on the Position of the Pole in the Scattering
Amplitude and on Its Residue}
\author{Yu.\,V.~Orlov}\email{orlov@srd.sinp.msu.ru}
\affiliation{Skobeltsyn Institute of Nuclear Physics, Moscow State
University, Moscow, 119991 Russia}
\author{B.\,F.~Irgaziev}\altaffiliation[Present address:\,]{GIK Institute of Engineering Sciences
and Technology, Topi, Pakistan} \affiliation{\,Institute of
Applied Physics, NUUZ, Tashkent, Uzbekistan}
\author{L.\,I.~Nikitina}
\affiliation{Skobeltsyn Institute of Nuclear Physics, Moscow State
University, Moscow, 119991 Russia}
\date{\today}

\begin{abstract}
Explicit expressions of the vertex constant for the  decay of a
nucleus into two charged particles for an arbitrary orbital
momentum  $l$ are derived for the standard expansion of the
effective-range function  $K_l(k^2)$, as well as when the function
$K_0(k^2)$ has a pole.  As physical examples, we consider the
bound state of the nucleus ${}^3\rm{He}$ and the resonant states
of the nuclei  ${^2}$He and ${^3}$He in the $s$-wave, and those of
${}^5\rm{He}$ and ${}^5\rm{Li}$ in the $p$-wave. For the systems
$Np$ and $Nd$ the pole trajectories are constructed in the complex
planes of the momentum and of the renormalized vertex constant.
They correspond to a transition from the resonance state to the
virtual state while the Coulomb forces gradually decrease to zero.
\end{abstract}

\maketitle

\section{INTRODUCTION}
The vertex constants (VC) $G_l^2$ ($l$ is the orbital momentum,
the other quantum numbers are omitted) for $a\leftrightarrow{b+c}$
virtual decay (synthesis) and the related residues of the partial
elastic scattering amplitude in the pole of a bound or resonance
state (when a decay is real) are considered to be independent
physical quantities. They play an important role in the theory of
nuclear reactions using Feynman diagrams and in astrophysics. For
the first time the vertex parts of diagrams, defining an input of
the specific mechanism into the cross-section of a direct nuclear
reaction, were considered in the frame of the dispersion
nonrelativistic approach proposed by I.\,S.~Shapiro more than 45
years ago \cite{Schapiro}. The main aim of this approach was
giving up the perturbation theory, on which the distorted wave
method is based, for processes with a strong interaction. The
selection criterion for the most important diagrams is  closeness
to the observable physical region of the singularity of the
corresponding amplitude along with the diagram ``weight'' produced
by the product of the vertex parts. Vertex constants are directly
related to the asymptotic normalization coefficient (ANC) of the
wave function, determining (along with the ``binding'' energy) the
system $a$ at large distances between particles $b$ and $c$. The
Coulomb repulsion effects must be taken into account for the
charged particles. In the present study the problem is solved in
the frame of an effective-range theory. The preliminary results
for the $s$-wave were published in our works \cite{Erem07,Orl09}.

As far as we know, this method of finding VC has not been
considered in detail in the literature with the exception of
\cite{Yarmukh07} where the authors attempt to find a relationship
between the vertex constant $G_l^{\rm{NC}}$ (the upper indexes
point to the nuclear and the Coulomb interactions) for the bound
state of the nucleus $a$ and the parameters of the standard
effective-range approximation for the elastic scattering of
charged particles $b$ with $c$ at arbitrary values of the orbital
angular momentum $l$. Besides the binding energy $\epsilon_{bc}$
of the nucleus $a$ the parameter list includes the ``scattering
length''  $a_l$ and the ``effective range''  $r_l$. (We put these
in quotes because the dimensions of $a_l$ and $r_l$ depend on $l$,
the length dimension this parameters have only at $l=0$).
Unfortunately, a serious error of fundamental importance was made
in [4]: Eq. (25) in \cite{Yarmukh07} (the numbering of the
formulas in this and in the next section corresponds strictly to
the numbering in \cite{Yarmukh07}), which relates the binding
energy to the ``scattering length'' and to the ``effective
range'', was written without allowance for the Coulomb interaction
(!). Equation (25) which is inappropriate in the case of charged
particles, was used there to derive  a formula for the
elastic-scattering amplitude \{see Eq. (23) in \cite{Yarmukh07}\},
and the correct expression for $\cot\delta^{\rm{NC}}_l(E)$ from
\cite{Mur83}, which takes into account both the Coulomb and
nuclear interactions, was substituted into it. As a result, the
``hybrid'' amplitude obtained thereby has neither a pole for the
$\epsilon_{bc}$ without the Coulomb interaction nor for the proper
binding energy value $\epsilon^{\rm{NC}}_{bc}$ with the Coulomb
interaction taken into account. Nevertheless the derivative of the
numerator was taken at the binding energy $E_{\rm{cm}} =
-\epsilon_{bc}$ (30) in order to obtain $G_l^{\rm{NC}}$. Thus the
expressions (30), (31) in \cite{Yarmukh07} for the vertex constant
of the virtual decay $a\rightarrow b+c$ into two charged fragments
are invalid. At first glance the situation can be remedied by
using the experimental value of the binding energy $\epsilon_{bc}$
in formula (31). But it is known that the parameters of the
effective-range function change when the Coulomb interaction is
taken into account. That is why one should not use formula (23)
for neutral particles, which gives $\epsilon_{bc}$ in terms of the
parameters of the effective-range theory for neutral particles to
find these parameters. Finding $a_l$ and $r_l$ by using data on
the vertex constant is just one of the objectives of
\cite{Yarmukh07}, this means using formulas (25) and (31) (or (34)
for VC). Invalid formulas (25) and (34) were applied in paragraph
4.2 in \cite{Yarmukh07} for the estimation of the scattering
lengths and corresponding effective radii for $\alpha t$
scattering in the $p$-wave by fitting the experimental ANC values
for the channel spin $S_B=$3/2 and 1/2. Because of the error
mentioned above the results of this paragraph are questionable.

We derive formulas for the renormalized scattering amplitude and
for the corresponding vertex constant in terms of the
effective-range approximation with the Coulomb interaction taken
into account. A more detailed analysis of the mistakes in
\cite{Yarmukh07} is given below.

The prime advantage of the approach proposed by us   is the
possibility of finding $G_l^2$  using  only experimental data,
which allows us to obtain  parameters of the effective-range
function $K_l(k^2)$. For the arbitrary $l$ values the standard
effective-range expansion has the following form (see, for
example, \cite{Mur83})
\begin{equation} \label{efradl}
K_l(k^2)= -1/a_l+{r_lk^2}/2 +\cdots,
\end{equation}
where the dimensions are as follows:  $[a_l]=L^{2l+1} ,[r_l] =
L^{-2l+1} (L$ -- is the length). The following notations are
adopted: $k = (2\mu E_{\rm{cm}})^{1/2}$  is the wave number, $\mu$
is the reduced mass, $E_{\rm{cm}}$ is the energy in the
center-of-mass (cm) frame. For brevity below we omit  the indexes
of the energy in the cm frame $E=E_{\rm{cm}}$. The energy in the
laboratory frame we denote as $E_{\rm{lab}}$. Here and below we
use the system of units where $\hbar$=$c$=1.

For the  $Nd$ system the effective-range function has the form
\begin{equation}\label{efrad}
K_0(k^2)=
(-1/a_0+C_2k^2+C_4k^4)/(1+k^2/{\kappa_0}^2).
\end{equation}
The pole of the effective-range function $K_0(k^2)$ in
(\ref{efrad}) at $ k^2=-{\kappa_0}^2$, i.e. at the energy
$E_{\rm{cm}}=- E_0$, where $E_0 = (3/4m){\kappa_0}^2$ (for the
$Nd$ system $\mu=(2/3)m$, $m$  is the nucleon mass), is a feature
of the doublet $Nd$ system \cite{Delv60,Oers67,Whit76,Sim87}. For
the $nd$ system, the pole is at a negative energy near the
threshold, i.e. $E_0>0$. For the $pd$ system the situation is much
more ambiguous. Due to the Coulomb repulsion, the measurements at
very low energies are complex and unreliable. As a result the
information on the scattering length and position of the pole of
$K_0(k^2)$ obtained in different works is ambiguous and
contradictory. In a few analyses of the phase shift it was found
that $E_0<0$. However, an analysis of the latest calculations
within the three-body problem involving three-particle interaction
gives $E_0>0$, with a pole located very close to the elastic
scattering threshold.

In the absence of the Coulomb interaction, the effective-range
function $\ K_l(k^2)$ is related to the phase shift $\delta_l(E)$
by the expression
\begin{equation}  \label{lneutral}
K_l(k^2)= k^{2l+1} {\cot \delta_l(E)}.
\end{equation}

In the presence of the Coulomb repulsion, the right-hand side of
the formula is transformed in the well-known way (see, for
example, \cite{VanHaer}) \footnote[1]{A misprint in the bracket
positions of this formula  was made in \cite{Orl09} but the right
formula was used for calculations.}:

\begin{equation} \label{CoulombKl}
K_l(k^2) = k^{2l+1} (c_{l\gamma})^{-1}\left[[2\pi \gamma
/(\exp(2\pi\gamma)-
 1)](\cot\delta_l^{\rm{C}}(E) -  i)+  2 \gamma  H(\gamma )\right],
\end{equation}
where
\begin{equation} \label{H-gamma-sigma}
H(\gamma)\equiv \Psi(i\gamma) + (2i\gamma )^{- 1} - \ln(i\gamma),
\end{equation}
\begin{equation} \label{cl-gamma}
(c_{l\gamma})^{-1}=\prod_{n=1}^l(1+\gamma^2/n^2),\quad
(c_{0\gamma})^{-1}=1,
\end{equation}
\begin{equation} \label{phase}
\delta^{\rm{C}}_l(E)=\delta_l(E)-\sigma_l(E).
\end{equation}

The following designations are used here: $\delta_l(E)$ is the
phase shift for the sum of the Coulomb and nuclear potentials,
$\sigma_l(E)$ is the Coulomb phase shift, determined by the
relation
\begin{equation}
\label{sigma} \exp(2i\sigma_l) =
{\Gamma(l+1+i\gamma)}/{\Gamma(l+1-i\gamma)},
\end{equation}
$\Psi(i\gamma)$ is the psi-function (logarithmic derivative of the
gamma-function), $\gamma =\lambda/k$ is the Sommerfeld Coulomb
parameter, $\lambda=\mu \alpha Z_b Z_c$, $\alpha=e^2/\hbar c$ is
the fine-structure constant; and $Z_I$ is the charge number of the
nucleus $I$. We  also use the Bohr radius $a_B=1/\lambda$. This
notation was accepted in \cite{Kok80}. Using the explicit
expression for $\Psi(i\gamma)$ in the form of infinite sum, we can
write $H(\gamma)$ as (see, for example, \cite{Arv74}):
\begin{equation} \label{sum}
H(\gamma)=\frac{i\pi}{\exp(2\pi\gamma)-
1}+\gamma^2\sum_{n=1}^\infty\frac{1}{n(n^2+\gamma^2)}-\ln(\gamma
)-\zeta,
\end{equation}
where $\zeta \approx$ 0.5772 is the Euler constant.

Kok \cite{Kok80} was the first to show that in the presence of the
Coulomb repulsion the scattering amplitude pole in the $s$-wave
for the virtual (antibound) state shifts from the negative
imaginary axis (Im$k<0$) to the fourth quadrant of the complex
plane and becomes a resonance pole at $k=k_{\rm{res}}=
{\rm{Re}}k_{\rm{res}}+ i {\rm{Im}}k_{\rm{res}}$. Since $H(\gamma)$
in (\ref{sum}) is an explicit function of the argument $ik$,
another pole arises at $k=-{\rm{Re}}k_{\rm{res}} + i
{\rm{Im}}k_{\rm{res}}$ in the third quadrant of the complex plane,
which is symmetric relative to the imaginary axis with respect to
$k_{\rm{res}}$. For the singlet $Np$  and the doublet $Nd$ systems
both poles are below the threshold (${\rm{Re}}E_{\rm{res}}<0$),
because $\mid {\rm{Im}}k_{\rm{res}}/{\rm{Re}}k_{\rm{res}}\mid>1$.
The mirror symmetry of zeros and poles with respect to the
imaginary momentum axis stems from the general symmetry properties
of the S-matrix (see, for example, \cite{Baz71,Sit75}).

The expression for $\cot \delta_0^{\rm{C}}(E)$, valid for real
positive energies $E$ in the physical region, is given in the
textbook by Landau and Lifshitz (Eq.(136.11) \cite{Landau63}),
where an expansion of type (1) is used.

For the sake of completeness, we present the corresponding formula
taken from \cite{Mur83}, which is valid at arbitrary $l$ value and
can be used for the analytical continuation into the complex
momentum plane.

\begin{equation} \label{CoulombMur}
K_l(k^2) = k^{2l+1}
(c_{l\gamma})^{-1}\left[[2\pi\gamma/[\exp(2\pi\gamma)-1]]
\cot\delta_l^C(E) +\gamma[\Psi (1+i\gamma )+\Psi(1-i\gamma
)-2\ln\gamma]\right].
\end{equation}

Eq.(\ref{CoulombMur}) is suitable for a parametrization of the
scattering phase shift. The corresponding expression for the
$p$-wave is given in the work \cite{Ahmed76}, where it was used
for finding the S-matrix poles for $N\alpha$ scattering near the
elastic threshold. We recalculate these pole positions more
accurately in the present work using the results of the phase
shift analyses of the experimental data known from the literature.
After that we calculate the renormalized VC, which were not
considered in \cite{Ahmed76}. They can be used for finding the ANC
of the corresponding Gamow wave functions.

In our work \cite{Erem07} we study the resonant subthreshold state
of the nucleus $^2$He and the resonance for the $\alpha\alpha$
system which has an extremely narrow width. The resonance
positions for these states were calculated by Kok \cite{Kok80}
using expansion (1). However, the vertex constants were not
considered. The parameters for the  $pp$ scattering were borrowed
from \cite{Brown}.

In  works \cite{Erem07,Orl09} we also studied the $^3$He bound
state and the subthreshold resonances in $pd$ scattering for the
most reasonable sets of constants in the expansion (2) which were
fitted recently in \cite{OrlIzv05,OrlOreNik02} using the doublet
phase shift for the $s$-wave which was calculated for few energy
values in the three-body approach \cite{Ki95,Ki96}, for the $NN$
potential AV18 and  the three-particle interaction UR-IX. The
necessity of using the calculation results for finding the
parameters of $K_0(k^2)$ is due to the fact that the experimental
data for the $pd$ scattering show large errors which increase
while the energy decreases, and because the results of the phase
shift analyses published in the literature are contradictory.

In the present work the trajectories for a transition from the
resonant subthreshold state to the virtual (antibound) state for
the $Np$ and the $Nd$ systems  were constructed with a gradual
decrease of the nucleon charge (the preliminary calculations were
fulfilled in \cite{Orl09}). As a result we demonstrate the general
physical nature of those corresponding states which differ only in
the Coulomb interaction. A comparison of the trajectories for the
different systems allows us to understand why the role of the
resonance (or the virtual level) in $Nd$ scattering is less
important than in the case of $Np$ scattering. The resonance
momentum trajectory at the transition from $pp$ to $np$ system was
previously calculated in  \cite{Csoto} using the
Eikemeier--Hakenbroich $NN$ potential, which is independent of
nucleon charge properties, and the Coulomb potential multiplied by
the coefficient $\xi$, which changes from 0 to 1. The analytical
continuation of the S-matrix onto the lower half-plane of the
complex momentum was fulfilled by the Schr\"odinger equation
solution. The resulting trajectory occurs close to the linear.

\section{METHOD FOR CALCULATING THE RENORMALIZED VERTEX CONSTANT}

The scattering amplitude for charged particles can be written as
the sum of the pure Coulomb amplitude $f_{\rm{Coul}}$(\textbf{k})
and the amplitude originated from a short-range (nuclear)
interaction in the presence of the Coulomb field (see, for
example, \cite{Gold})
\begin{equation} \label{CoulombNucl}
f({\bf k})= f_{\rm{Coul}}({\bf k})+\it f_{\rm{Nucl}}({\bf k}),
\end{equation}
\begin{equation} \label{Coulomb}
f_{\rm{Coul}}({\bf k}) =
\sum_{l=0}^\infty\frac{2l+1}{2ik}(\exp(2i\sigma_l)-1)
P_l(\cos{\theta}),
\end{equation}
\begin{equation} \label{Nucl}
f_{\rm{Nucl}}({\bf k}) =
\sum_{l=0}^\infty\frac{2l+1}{2ik}\exp(2i\sigma_{l})
(\exp(2i\delta_{l}^{\rm{C}})-1){P_{l}(\cos\theta)}.
\end{equation}

Equation (\ref{Nucl}) is written for spinless particles. The
generalization to the case of spins taken into account is
elementary because the Coulomb interaction does not depend on the
spin values. We consider  amplitudes which are diagonal relative
to spin and to orbital angular momentum values. A cross-section
includes non-diagonal amplitudes as well when the spins are taken
into account. In any case VC depends on one set of the quantum
numbers $l$ and $s$.

The Coulomb-nuclear partial scattering amplitude with the orbital
angular momentum $l$, which has the poles we take interest in, can
be written as (see (\ref{Nucl})) \footnote[2] {The obviously
important  well known factor $\exp(2i\sigma_l)$  written in the
equation below  was not included in  \cite{Yarmukh07}.}:

\begin{equation} \label{fC0}
f_{l}^{\rm{C}} = \exp(2i\sigma_l) f_l,
\end{equation}
\begin{equation} \label{f0}
f_{l} = (\exp(2i\delta_l^{\rm C}-1)/2ik =1/(k \cot\delta_l^{\rm C}
-ik).
\end{equation}

Let us find $(k\cot\delta_l^{\rm C}-ik)$ from (\ref{CoulombKl})
and insert the result into (\ref{fC0}), (\ref{f0}). We obtain the
following formulas:

\begin{equation} \label{f0K}
f_l^{\rm{C}} = \tilde{f}_{lN}^{\rm{C}} k^{2l}\phi_l(k),
\end{equation}
where
\begin{equation} \label{f0C}
\tilde{f}_{lN}^{\rm{C}} = [K_l(k^2)-2k^{2l+1}
({c_{l\gamma}})^{-1}\gamma H(\gamma)]^{-1},
\end{equation}
\begin{equation} \label{phi-l-k}
\phi_l(k)=\exp(2i\sigma_l){\left[2\pi\gamma/[\exp(2\pi\gamma)-1]\right]}(c_{l\gamma})^{-1}.
\end{equation}

Let us simplify the expression for  $\phi_l(k)$. After simple
transformations using the relationship (see (3.2) on page 54 in
\cite{Janke})

\begin{equation} \label{JankeP}
\Gamma(l+1+i\gamma)\Gamma(l+1-i\gamma)=\frac{\pi
P_{l+1}(i\gamma)}{\gamma \rm{sh}(\pi\gamma)},
\end{equation}
where $P_n$ (not the Legendre polynomial) is defined by formulas

\begin{equation} \label{f0C-gamma}
P_1(i\gamma) =\gamma^2,\,\,\,\, P_{l+1}(i\gamma) =\gamma^2
\prod_{n=1}^l(n^2+\gamma^2),
\end{equation}
we receive the expression
\begin{equation} \label{phi-lkshort}
\phi_l(k)=\left(\frac{\Gamma(l+1+i\gamma)}{l!}\right)^2
e^{-\pi\gamma}.
\end{equation}

The function $\phi_l(k)$, which does not contain  information
about the nuclear interaction, coincides with the factor at the
renormalized scattering amplitude given in the review
\cite{Blokh84} (see Eq.(3) in \cite{Blokh84}).

Correspondingly, we can write the analog of the renormalized
partial amplitude for a particle  scattering by a short-range
potential in the presence of the Coulomb interaction in terms of
the effective-range function $K_l(k^2)$ as [in the following, we
use a tilde sign $(\sim)$ above the designation of  renormalized
quantities]

\begin{equation} \label{Q0C}
\tilde{f}_{lN}(k) =\tilde{f}_{lN}^{\rm C} k^{2l}=\frac{k^{2l}}
{\left[K_l(k^2)-2 \lambda k^{2l}
H(\gamma)(c_{l\gamma})^{-1}\right]}.
\end{equation}
According to this formula, it is obvious that the zero of the
denominator in (\ref{Q0C}) when $k=p$ is an amplitude pole.

For the standard effective-range expansion the pole position is
derived from the equation (with (\ref{efradl}) taken into account)
\begin{equation} \label{poleCoulomb}
-1/a_l+{r_lp^2}/2 +\cdots= 2\lambda p^{2l}
H(\lambda/p)[(c_{l\gamma})^{-1}]|_{k=p}.
\end{equation}

In the absence of the Coulomb interaction the pole position is
defined by formula (\ref{lneutral}) at $\it{\cot}\delta_l^{\rm
C}=i$ (see (15)), which leads to the following simple equation for
the pole position in the effective-range approach with taking the
first two  terms in (\ref{efradl}):
\begin{equation} \label{poleWithoutCoulomb}
K_l(p^2)= -1/a_l+{r_lp^2}/2=ip^{2l+1}.
\end{equation}
For the bound state $a$ of two  particles when one of them is
chargeless this implies the equation ($p=i\kappa,
\kappa=\sqrt{2\mu \epsilon_{bc}}>0, \epsilon_{bc}$ is the binding
energy)
\begin{equation} \label{boundWithoutCoulomb}
1/a_l+{r_l\kappa^2}/2=(-1)^l \kappa^{2l+1},
\end{equation}
which differs essentially from the equation (\ref{poleCoulomb})
for charged particles.

Nevertheless, just  equation (\ref{boundWithoutCoulomb}) was used
(see (25) in \cite{Yarmukh07}) where in addition the factor
$(-1)^l$ was lost. Correspondingly, equation (24) in
\cite{Yarmukh07} is also incorrect where  formula (25) was used to
write  the scattering amplitude. In our paper we use the right
equation (\ref{poleCoulomb}) from which the main equation of
\cite{Kok80} was also derived for determining amplitude poles in
the momentum complex plane, that is, on the nonphysical sheet of
energy. The authors of \cite{Yarmukh07} do not discuss the
introduction of the renormalized VC which is real for a bound
state (see \cite{Blokh84}). In particular, the factor
$\exp(2i\sigma_l)$ was not included in their formula (19) for the
Coulomb-nuclear amplitude. We note that it is precisely the
renormalized VC is related to the ANC by the simple relationship
(see \cite{Blokh77}).

Using the VC definitions which is well known from \cite{Blokh77}
we obtain the following expression for the renormalized VC:
\begin{equation} \label{Gren2}
\tilde{G}_l^2 = -(\pi/\mu^2) \lim_{k\to{i\kappa}}(k^2 +
\kappa^2)\tilde{f}_{lN}(k).
\end{equation}
The equation (\ref{Gren2})  can be rewritten as
\begin{equation} \label{Gren2dif}
\tilde{G}_l^2 =
\frac{(-2\pi/\mu^2)p^{2l+1}}{\frac{d}{dk}\left[K_l(k^2)-2 \lambda
k^{2l} H(\gamma)(c_{l\gamma})^{-1}\right]_{k=p}},
\end{equation}
where for the standard expansion (\ref{efradl})
$\frac{d}{dk}[K_l(k^2)]_{k=p}=r_lp+\cdots$,\,\,\,\, $p=i\kappa$ is
the position of the pole amplitude for a bound   ($\kappa
>0$ is real)or a resonance state (Im$p<0$, Re$\kappa<0$).

We considered in \cite{Erem07} both the conventional expansion of
the effective-range function $K_0(k^2)$ in powers of $k^2$ ($a_0$
is the scattering length, $r_0$ is the effective radius, $P$, $Q$
are the shape  parameters),
\begin{equation} \label{efrad0}
K_0(k^2)= -1/a_0+{r_0k^2}/2-P{r_0}^3k^4+Q{r_0}^5k^6-\cdots,
\end{equation}
which leads to
\begin{equation} \label{Gren2(1)}
{{\tilde{G}_0}^2} =
\frac{2\pi\kappa/\mu^2}{\varphi(x)-[r_0\kappa+4P(r_0\kappa)^3 +
6Q(r_0\kappa)^5 +\cdots]},
\end{equation}
and the function with a pole (\ref{efrad}) ($C_0=-1/a_0$), when we
obtain \footnote[3]{Here, the misprint made in \cite{Erem07} is
corrected: the factor $\kappa_0^2$ is added in the denominator in
front of the  square bracket.}

\begin{equation} \label{Gren2(2)}
{\tilde{G}_0}^2 = \frac{2\pi\kappa/\mu^2}{\varphi(x)-{2\kappa
\kappa_0^2}
[-C_0+C_2\kappa_0^2-C_4\kappa^2({2\kappa_0^2}-\kappa^2)]/{(\kappa_0^2-\kappa^2)^2}}.
\end{equation}
The function $\varphi(x)$ has the form
($x=\lambda/\kappa=1/a_B\kappa$)
\begin{equation} \label{phi}
\varphi(x)=-1-2x+2x^2\Psi'(x).
\end{equation}

Let us remember that due to (15) the positions of the poles
($k_{\rm{res}}=i\kappa$) of the scattering amplitude $f(k)$ for
the bound or resonance (or virtual) states are determined by the
condition
\begin{equation} \label{pole}
\cot\delta_l^{\rm{C}}(E)=i.
\end{equation}

\section{VERTEX CONSTANTS FOR PARTICULAR NUCLEI}

The case of the $s$-wave $(l=0)$ was considered in our papers
\cite{Erem07,Orl09}, where all the necessary formulas were given
as well as the numerical calculation results for the nuclear
systems $NN, Nd$ and $\alpha\alpha$. Several results from
\cite{Erem07,Orl09}  are given below for the sake of completeness.
In the present paper we study more thoroughly the trajectories of
the resonance momentum and the renormalized VC while the charge of
the one of the particles gradually goes to zero.

\subsection{The proton-proton system. The ground state of the
$^2{\rm{He}}$ nucleus}

The results of the transcendental equation solution
(\ref{poleCoulomb}) with the parameters set, taken from
\cite{Brown} (without taking inaccuracy into account), found in
\cite{Erem07}, are $k_{ ^2{\rm{He}}}=(0.0644-i0.0871)$ fm$^{-1}$,
$ E_{^2{\rm{He}}}=(-142-i465)$ keV. They differ only in the last
figure from those obtained in \cite{Kok80}: $ k_{
^2{\rm{He}}}=(0.0647-i0.0870)$  fm$^{-1}$,$\quad E_{
^2{\rm{He}}}=(-140-i467)$ keV. In \cite{Erem07}  for the first
time   the renormalized vertex constant squared was also found
${\tilde{G}}^2=-(0.060+i0.051) $fm. The calculation results change
only weakly if the shape parameters $P=Q=0$.

Let us now consider the transition from the $pp$ system to a
system without Coulomb interaction. Either $nn$ or  $np$ system
can be taken from the isotopic triplet. For the sake of certainty,
we choose the singlet $np$ system, for which (in the approximation
$P=Q=0$)  $a_0^{np}=(-23.719\pm0.013)$ fm, $\quad
r_0^{np}=(2.76\pm0.05)$ fm (see \cite{Brown}). With this set of
parameters, the virtual level energy $E_v=-66$ keV.

In \cite{Erem07} the $pp\rightarrow np$ transition was set by the
linear relations
\begin{equation} \label{xieffrada}
a_0(\xi)=a_0^{pp}\xi+a_0^{np}(1-\xi);
\end{equation}
\begin{equation} \label{xieffrad}
r_0(\xi)=r_0^{pp}\xi+r_0^{np}(1-\xi).
\end{equation}
This approximation leads to a highly nonlinear trajectory of the
resonance pole $k_{\rm{res}}(\xi)$ in the momentum complex plane,
in contrast to the result of \cite{Csoto}. It is known  (see, for
example, \cite{Brown}), that the effective radius only  changes
slightly  when the Coulomb interaction is taken into account;
therefore (\ref{xieffrad}) is a good approximation. The decisive
factor is the dependence of the scattering length on the product
of charges, which significantly differs from the linear.

A monograph by Brown and Jackson \cite{Brown}   gives an
approximate formula for the relation between the  $pp$ and $np$
scattering lengths derived in accordance with  the calculations
performed long ago by Landau and Smorodinsky (see (4.29) and
reference in  \cite{Brown}).  In \cite{Brown} it is noted, that,
in spite of crude  approximations, this formula  with an accuracy
within a few percent is valid for the potentials, which permit us
to describe well the $pp$ scattering data. Analogous formula had
been derived in the work by Schwinger  \cite{Schwinger50}.

Introducing the coefficient  $\xi$ into this formula by the
replacement $\lambda\rightarrow\xi\lambda$,  we obtain the
following relation, determining the dependence  of  the scattering
length $a_p(\xi)$ on $\xi$ \cite{Erem07}:
\begin{equation} \label{Brown}
1/a_p(\xi)=1/a_0^{np}+2\lambda\xi\:(\ln(2R\lambda\xi)+0.33).
\end{equation}

\begin{figure}[bp]
\resizebox*{0.47\textwidth}{!}{\includegraphics{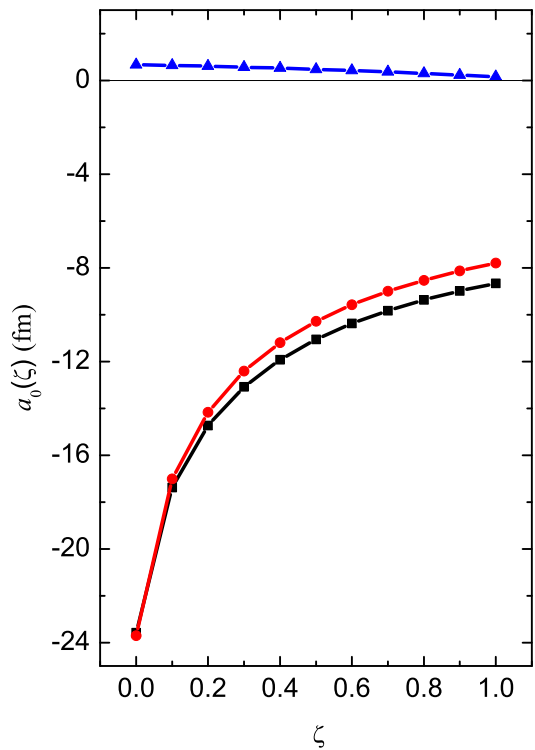}}
\caption{The $a_0(\xi)$ dependencies of the singlet $Np$
scattering length (two bottom curves) and of the doublet $Nd$
scattering length (top curve, triangles)  on the coefficient $\xi$
(multiplier before of the Coulomb potential). The results of the
$a_0(\xi)$ calculations are obtained using (35) (following the
Landau--Smorodinsky--Schwinger relation between $a_0^{pp}$ and
$a_0^{np}$) (circles) and  by solving the Schr\"odinger equation
with the Yukawa potential (squares).} \label{fig1}
\end{figure}
The value of $R$ was not found in \cite{Brown}. It was only noted
that the dependence on this parameter should be weak because $R$
is under the logarithm sign. Solving the transcendental  equation
(\ref{Brown}) at $\xi=1$ we find $R=1.74$ fm. This value is within
the margins of the radius spread for different potentials with two
parameters given in \cite{Brown} for the singlet $np$ system. The
dependence in the form (\ref{Brown}) is approximated well by the
relation \cite{Erem07}
\begin{equation} \label{Inverse}
1/a_p(\xi)=(1-\xi^\beta)/a_0^{np}+\xi^\beta/a_0^{pp}
\end{equation}
at $\beta=0.7$. Clearly, nonlinear dependence $a_p(\xi)$ is shown
in Fig.\ref{fig1}.

Finding out the dependence of the parameters of the
effective-range function on  $\xi$ is a dynamic problem. Let us
introduce into the Schr\"odinger equation an interaction $\xi
V_{\rm{C}}(r)$, instead of the Coulomb potential $V_{\rm{C}}(r)$,
where the coefficient $\xi$ ranges within  $0\leq \xi\leq 1$.
Thus, one must replace the Sommerfeld parameter
$\lambda\rightarrow\xi\lambda$ in all formulas. We obtain the
results shown in Fig. \ref{fig1} by solving the Schr\"odinger
equation in the continuum for the $Np$ and $Nd$ scattering to find
the energy dependence of the effective-range functions and their
parameters. We use the Yukawa potential
\begin{equation} \label{Yukawa}
V(r)=-V_0 (R/r)\exp(-r/R),
\end{equation}
as a nuclear interaction for which the parameters of the
neutron-proton interaction in the singlet state are given in
\cite{Brown}.

\begin{figure}[bp]
\resizebox*{0.7\textwidth}{!}{\includegraphics{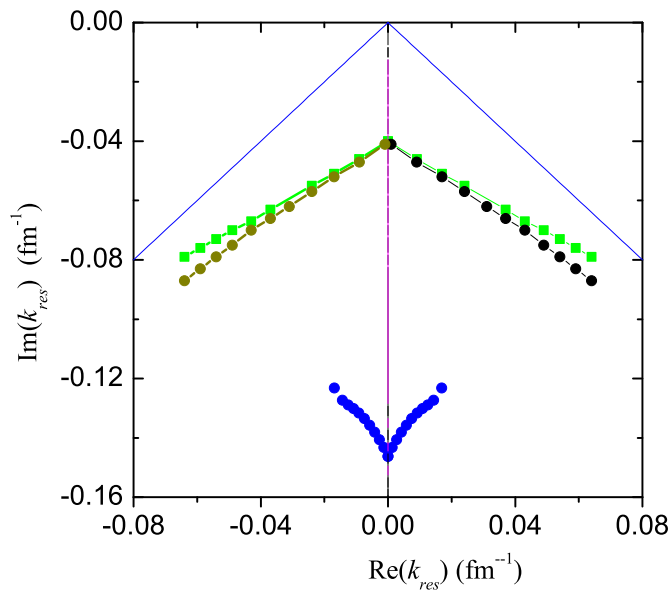}}
\caption{Trajectories of $k_{\rm{res}}$ for the transition from
the resonance pole to the virtual (antibound) state at a gradual
decrease in the Coulomb interaction to zero. The top curves are
trajectories for the $Np$ system. The points correspond to
different $\xi$ values in the interval 0--1. The circles present
the calculations, obtained by formula (\ref{Brown}); the squares
for the $Np$ and points (bottom curves) for the $Nd$ systems are
obtained by solving the Schr\"odinger equation with the Yukawa
potential. The thin straight lines are the angle bisectors of the
third and fourth quadrants of the complex plane.} \label{fig2}
\end{figure}

Let us now construct the trajectories for $k_{\rm{res}}(\xi)$ and
${\tilde{G}}^2(\xi)$. We find the parameters of the
effective-range function  by fitting  the phase shift values
calculated at low energies where the effective-range expansion is
valid. The corresponding pole trajectories in the complex momentum
plane are given in Fig. \ref{fig2}. They are close to the linear
as in \cite{Csoto}.

The calculation results show that, in accordance with the
conclusion of \cite{Brown}, it is only the dependence of the
scattering length $a_p(\xi)$ on $\xi$ that is  important. The
effective radius changes only slightly and can be described by the
linear dependence (\ref{xieffrad}). The corresponding curve is
situated a little lower than the curve  following from the
Landau--Smorodinsky--Schwinger (\ref{Brown}) (or (\ref{Inverse})
at $\beta=0.7$), because in this case the $pp$ scattering length
differs from the experimental one.

\begin{figure}[bp]
\resizebox*{0.9\textwidth}{!}{\includegraphics{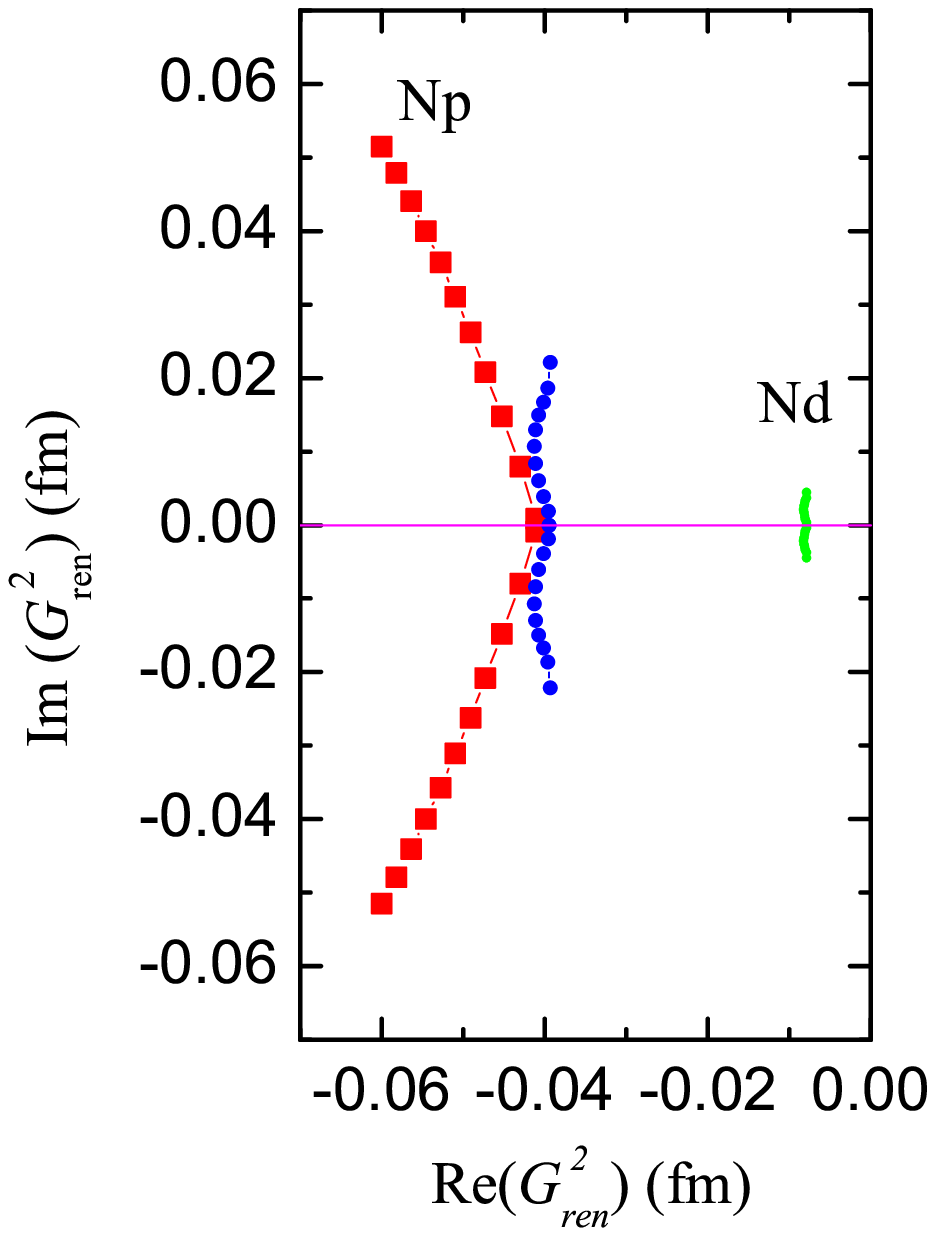}}
\caption{Trajectories as in Fig. \ref{fig2}, for the square of the
renormalized vertex constant, $G_{\rm{ren}}^2\equiv(\tilde{G})^2$,
calculated with the Yukawa potential for the $Np$ (squares) and
$Nd$ systems (on the left and right, respectively); the trajectory
5($\tilde{G})^2$ is also shown for the $Nd$ system.} \label{fig3}
\end{figure}

The trajectories ${\tilde{G}}^2(\xi)$ in the complex plane of the
renormalized vertex constant for the $Np$ system given in Fig.
\ref{fig3} are also close to the linear. They are symmetric with
respect to the real axis -- that is, a complex conjugate. The
contribution of the poles, located symmetrically with respect to
the imaginary momentum axis, to the partial amplitude on the real
energy axis is real in this case, as it should be for real
interaction potentials.

\subsection{The doublet state of the $pd$ system in the ${s}$-wave}

With the strong correlations between the parameters  $C_2, C_4,
E_0$ â (\ref{efrad}) and the scattering length $a_0$ taken into
account (see, for example, \cite{Sim87,OrlOreNik02}) it is
suitable to write down the effective-range function in the form
\begin{equation}
\label{efradE}
K_0(k^2)=\frac
{(-1+c_2k^2+c_4k^4)}{(a_0+d_2k^2)}=\frac{-1+b_1E+b_2E^2}{(a_0+d_1E)}.
\end{equation}
The parameters which appear in this formula have a much weaker
dependence on $a_0$. An approximation of the results of three-body
calculation of the phase shifts in the explicit form (\ref{efrad})
and (\ref{efradE}) can be considered as an analytical continuation
of the partial scattering amplitude onto complex energy $E$
values. This allows us to find the whole set of the low energy
characteristics for the $Nd$ system. The set includes the
positions of the scattering amplitude poles for the bound and for
the resonance (or virtual) state which are defined by zeroes of
the denominator of (22) due to the condition (\ref{pole}), along
with the pole position ($-E_0$) of the function $K(k^2)$.

The corresponding residues at the poles are connected with the
nuclear vertex constants $G_t^2$ and ${\tilde{G}}_\tau^2$ (see
\cite{Blokh84,Blokh77}) or with the asymptotic normalized
coefficients $C_t^2$ and $C_\tau^2$ (see, for example,
\cite{Friar82}) for the nuclei $^3\rm{H}$ and $^3\rm{He}$, which
for the sake of brevity are denoted by the letters $t$ and $\tau$
correspondingly. As  is noted in the introduction, the available
phase shift analyses of the experimental data for the $pd$
scattering do not give an unambiguous result. In the first place,
this affects the scattering length. Because of this, the analyses
of both the modern three-body calculations with three-particle
forces taken into account \cite{Ki95,Ki96,Ki97} and the
calculation results in the frame of $N/D$ method \cite{Saf89} were
fulfilled in \cite{OrlIzv05,OrlOre06}.

Different  physical observables were used for the parameter
fitting: the $s$-wave doublet phase shift for the $pd$ scattering
at the smallest energy values considered in the literature, the
binding energy of $^3$He, and the position of the subthreshold
resonant pole, corresponding to the virtual state of $^3$H,  which
was calculated previously by the $N/D$ method in \cite{Saf89},
where the residue of the scattering amplitude was calculated as
well. The different fitting variants were given in the tables of
\cite{OrlIzv05,OrlOre06}.

The renormalized VC values for the selected fitting variants are
given in Table 1 of our work \cite{Erem07} where the right
$E_{\rm{res}}$ values and the parameters of the effective-range
function were presented as well \footnote[4]{The accurate
$E_{\rm{res}}$ results   for the variants No~3, 4 are given in
\cite{OrlOre06} (in \cite{OrlIzv05} the values of $k_{\rm{res}}$
were  given by mistake instead of $E_{\rm{res}}$ for these
variants). Besides, for the case of variant No~14 in
\cite{OrlOre06} the values of  $a_{pd}C_2$, $a_{pd}C_4$ are wrong,
they should be replaced by the values given for this variant in
the table of the paper \cite{OrlIzv05} (the lower line).}. In the
second block of this table the variants for the Argonne $NN$
potential (AV18) with three-nucleon forces taken into account in
the Urbana form (UR-IX) (see References in \cite{Ki95,Ki96})  are
considered for the value  $^2a_{pd}$ = 0.024 fm, obtained in
\cite{Ki97} published later where the pole position of the
effective-range function is also found ($E_0$ = 3.13 MeV). This
simplifies the fitting of the remaining parameters. In this case
it is possible to reproduce precisely the scattering length value.
Meanwhile the approximate phase shift behavior at low energy well
describes  the original three-body results for the interaction
AV18+(UR-IX) no matter whether the binding energy is fitted or
not. For this reason, we consider  $^2a_{pd}$ = 0.024 fm as the
most reliable theoretical estimation to date.

The  resulting  parameter value $c_2={^2a_{pd}}C_2=-58.1$ fm$^2$
found by us can be compared with the result of the analysis of the
phase shift calculated in the energy region $E_{\rm{cm}}<450$ keV
by the Pisa group (see \cite{Ki97})  using the formula
(\ref{efradE}) when $c_4=0$. The result obtained which corresponds
to the value $c_2$=$-56.7$ fm$^2$  agrees well with the results of
the analysis in \cite{OrlOre06}, especially for the variant No~10
($c_2=-56.6$ fm$^2$) (see Table 1 in \cite{Erem07}).

\subsubsection{ \textbf{The bound state $^3${\rm{He}} characteristics}}

The renormalized VC for the $^3$He bound state calculated in
\cite{Erem07} in the effective-range approximation (2) occurs to
be greatly underestimated (by about a factor of 2) in comparison
with the results of both experimental analyses and theoretical
calculations. This is due to the fact that the $^3$He bound state
pole is situated far away from the convergence region for the
expansion of the numerator (\ref{efrad}) in the powers of $k^2$.

Unfortunately, the majority of available publications on the VC
for $^3$H and $^3$He are some 20 years old. In addition, different
papers report values of either the renormalized VC ${\tilde{G}}^2$
or the asymptotic normalization constant (ANC) $C^2$. Table 1 in
\cite{Erem07} presents both the renormalized VC and the ANC for
the $^3$He bound state. The following relation between the VC and
the ANC (see, for example,  [28, 29]) is used:

\begin{equation} \label{G2toANC}
{{\tilde{G}_\tau}^2}=(27/4)\pi\kappa_\tau{\lambda_N}^2 {C_\tau}^2,
\end{equation}
where  $\lambda_N$=$\hbar/mc$, $\kappa_\tau$ is the wave number of
the bound $^3$He nucleus. This relation corresponds to the
definition of the asymptotic behavior of the normalized-to-unity
relative motion wave function of the proton and the deuteron in
the form (for the $s$-wave)
\begin{equation} \label{asymp}
\psi(r)\rightarrow C_\tau  N_{ZR}\left[
W_{-\gamma,1/2}(2\kappa_\tau r)\right]/r, \qquad  {r\to\infty},
\qquad N_{ZR}= (2\kappa_\tau)^{1/2},
\end{equation}
where $r$ is the relative distance (Jacobi coordinate) between the
nucleon and the center of mass of the deuteron. This definition of
the ANC corresponds to the value $C_\tau$=1 in the zero range
approximation, when the wave function coincides with its
asymptotic form (\ref{asymp}) for any ranges down to $r=0$.

For comparison, we give the VC and ANC data from \cite{Friar82},
where the ANC is introduced by the relation
\begin{equation} \label{asympFriar}
\psi(r)\rightarrow C_\tau N_W [W_{-\gamma,1/2}(2\kappa_\tau r)]/r,
\qquad {r\to\infty},
\end{equation}
where \footnote[5]{There is a misprint in the analogous formula
(38) in  \cite{Erem07}: the exponent of power  in $(N_W)^{-2}$ is
absent.}
\begin{equation} \label{NW}
{(N_W)}^{-2}=\int\limits_0^\infty\ [W_{-\gamma,1/2}(2\kappa_\tau
r)]^2 dr.
\end{equation}

The analytical expression for  $N_W$ is given in \cite{Friar82},
where $N_W/N_{ZR}$ = 1.024  for $^3$He (see \cite{Friar82}). This
factor is taken into account below in the calculation of the VC
from the results for $C_\tau^2$ from (\ref{asympFriar}). The
calculations in  \cite{Friar82} were carried out by solving
Faddeev type equations in the configuration space for the
Malfliet--Tjon (I--III) (MT)  and Reid soft-core (RSC) $NN$
potentials. The results are as follows ($\epsilon_\tau$ is the
binding energy of the $^3$He nucleus  in MeV, $\tilde{G}_{\tau}^2$
in fm):
\begin{equation} \label{MTresults}
{(\rm{MT})} \qquad\epsilon_\tau = 7.87;\quad C_\tau^2=3.90;\quad
\tilde{G}_\tau^2= 1.63;
\end{equation}
\begin{equation} \label{RSCresults}
{(\rm{RSC})} \qquad\epsilon_\tau = 6.39 ;\quad C_\tau^2=3.14;\quad
\tilde{G}_\tau^2= 1.12.
\end{equation}
Coulomb effects are calculated in (\ref{MTresults}) and
(\ref{RSCresults}) within the point charge approximation.
Allowance for spread of the charge does not affect the VC and ANC
values to a  precision of one-hundredth. The results for the RSC
potential are considerably underestimated in comparison with those
for the MT potential, first of all because the nucleus is
underbound for the RSC potential. In \cite{Ki97} the binding
energies of the $^3$H and $^3$He nuclei which are near the
experimental values were found for the   $NN$ potential AV18 with
the three-particle forces UR-IX taken into account. We give below
the results (the units are the same as in (\ref{MTresults}) and
(\ref{RSCresults})) for  $^3$H and $^3$He with the interaction
AV18+(UR-IX):
\begin{equation}\label{AV18URIXresults}
\qquad\epsilon_t= 8.49;\quad C_t^2=3.44;\quad G_t^2= 1.44; \qquad
\qquad\epsilon_\tau = 7.75;\quad C_\tau^2=3.53; \quad
\tilde{G}_{\tau}^2= 1.39.
\end{equation}

It has long been established in the literature that there is a
correlation between the binding energy and the vertex constant
(see, for example, \cite{OrlOreNik02} and references therein).
Therefore, a good reproduction of the experimental $\epsilon_\tau$
value is important for obtaining a reliable value of the VC. The
two-body potential model \cite{Petr88,Tom87}, where the binding
energy and the scattering length are adjustable parameters, meets
this condition.

Recently we have carried out the corresponding calculations for
various potentials \cite{Irg06}, where the potential parameters
were given. The parameters of the  Hulth\'{e}n and Yukawa
effective $nd$ potentials were  fitted to the experimental values
of the $^3$H binding energy and to the doublet $nd$ scattering
length $^2a_{nd}$. On the assumption of charge independence, the
same nuclear potential was used for the doublet $pd$ system. In
addition, calculations were carried out for the potentials
proposed in paper by Tomio et al. \cite{Tom87} (versions A and B).
In the case of version B, we refined the parameters beforehand to
get the correct doublet $nd$ scattering length (see reference in
\cite{Irg06}).

The results of the calculations are presented in Table
\ref{table1} \footnote[6]{For the Hulth\'{e}n potential in Table
\ref{table1} of the present work we  give more accurate
calculation results of the ANC and VC for  $^3$H and $^3$He.}.  It
is noteworthy that $G_t^2$ and  $\tilde{G}_\tau^2$ are close in
value despite the noticeable difference of the ANCs. The Coulomb
difference of the $^3$He and $^3$H  binding energies ($\approx$
0.7 MeV) is in rather good agreement with the experimental value
$\Delta E_c$=0.76 MeV.

\begin{table}
\caption{The nucleon separation energies $B_t$ and $B_\tau$ (in
MeV) for the nuclei $^3$H and $^3$He, the vertex constants
$G_{t}^2$ and renormalized $\tilde{G}_\tau^2$ (in fm), the
asymptotic normalized coefficients squared $C_t^2$ and $C_\tau^2$,
calculated in the two-body model for the different effective $nd$
potentials (for the Hulth\'{e}n potential the values are more
accurate than  in comparison with those given in
\cite{Erem07}).\label{table1}}.
\begin{ruledtabular}
\begin{tabular}{|l|ccccccc|}
 \,\,Potential&$B_t$&$C_t^2$&$G_t^2$&$B_\tau$&$C_\tau^2$&$\tilde{G}_\tau^2$&$^2a_{nd}\quad$
\\&&&&&&&\\
 \hline \,\,Hulth\'en&6.26&3.47&1.46&5.57&3.68&1.46&0.65\\
\,\,Yukawa&6.26&3.06&1.28&5.64&3.28&1.31&0.66\\
\,\,Tomio et al. (B)&6.26&3.56&1.49&5.60&3.77&1.50&0.65\\
\,\,Tomio et al.
(A)&6.26&3.94&1.65&5.58&4.18&1.66&0.68\\
\end{tabular}
\end{ruledtabular}
\end{table}

The results for  $^3$He agree with the conclusion in
\cite{OrlIzv05,OrlOre06,Orl06} concerning the convergence domain
of expansion (2) limited by the energies $|E_{\rm{cm}}|\leq 0.74$
MeV. This domain is determined by the position of the nearby
singularity  of the partial wave amplitude for the $pd$ scattering
at the energy $E_{\rm{cm}}=-\epsilon_d/3$  ($\epsilon_d$ is the
deuteron binding energy) corresponding to  the Feynman diagram for
the one-nucleon exchange.  The poles for the bound states of the
$Nd$ system turn out to be  far beyond the convergence domain. At
the same time the resonance of $^3$He lies within the convergence
domain as the pole for the virtual triton  and therefore   its
position and characteristics should be less sensitive to   the
variant of fitting and thus more reliable.   We think that the
best are the fits  No~7, 8 in \cite{OrlOre06}.

\subsubsection{\textbf{Subthreshold resonance of the $pd$ system}}

Our calculation of the energy of the subthreshold resonance for
the the parameter set of the effective-range function taken from
\cite{Ki97} leads to the value $E_{\rm{res}}^{pd}=-(0.315+i0.102)$
MeV, which differs only in the third decimal figure from the
corresponding value $E_{\rm{res}}^{pd}=-(0.319+i0.099)$ MeV  for
the variant No~8 in Table 1 of \cite{Erem07}, where the binding
energy of $^3$He serves as an additional fit value. Let us compare
our result $E_{\rm{res}}^{pd}\approx-(0.32+i0.10)$ MeV with other
results published in the literature.

In \cite{Saf89}  the value $E_{\rm{res}}^{pd}=-(0.432+i0.032)$ MeV
was obtained by the $N/D$ method. The absolute value of the real
part is slightly overestimated in the $N/D$ method whereas the
absolute value of the imaginary part is smaller by about a factor
of three. In  \cite{Csoto} the value
$E_{\rm{res}}^{pd}=-(0.432+i0.56)$ MeV is given, which was
obtained by solving the Faddeev equation for the
Eikemeier--Hackenbroich $NN$ potential. Its real part coincides
with the result of \cite{Saf89}, but the absolute value of the
imaginary part is larger than in \cite{OrlOre06} by about a factor
of five and is an order of magnitude larger than in the $N/D$
method \cite{Saf89}, which calls for explanation. Note that the
position of the virtual pole of the triton found in \cite{Csoto}
($B_v$ = 1.62 MeV) is also considerably  larger (by a factor of
more than 3) than other  estimates reported in the literature.

Let us now construct the trajectories of the position of the pole
and residue in it at the transition from $pd$ to $nd$. As in the
case of the nucleon-nucleon system, we use the Yukawa-potential
model (\ref{Yukawa}) with the parameters and the correponding
values of the physical quantities ANC and VC calculated by solving
the Schrodinger equation for the bound  and continuum states
(lengths in fm, energies in MeV):
\begin{equation} \label{ndYukawaParameters}
{(\rm{Yu}, {nd})} \qquad  R=2.77, \qquad V_ 0=17.46; \qquad
C^2_t=3.06,\qquad \tilde{ G}^2_t=1.28.
\end{equation}
The parameters in (\ref{ndYukawaParameters}) are fitted to the
triton binding energy ($\epsilon_t=8.48$ MeV) and to the doublet
scattering length ($^2a_{nd} = 0.65$ fm) for the $nd$ scattering
in the $s$-wave. This model describes well the energy behavior of
the phase shift for the doublet  $Nd$ scattering  \cite{Irg06} and
other  low-energy characteristics which were discussed above.

We introduce the coefficient $\xi$ into the product of charges and
find the dependence of the $s$-wave phase shift $\delta_0(E)$ on
$E$ in the range of the convergence of the effective-range
function $K_0^{(\xi)}(k^2)$  at different $\xi$ values  $(0\leq
\xi\leq 1)$.

\begin{figure}[bp]
\resizebox*{0.5\textwidth}{!}{\includegraphics{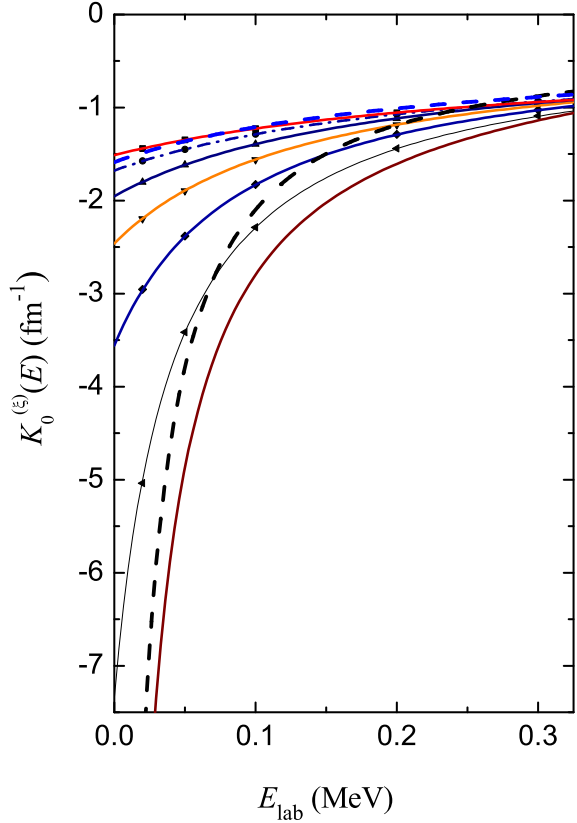}}
\caption{The dependence of the doublet effective-range function
$K_0^{(\xi)}(E)$ on the energy $E_{\rm{lab}}$ for the $Nd$
scattering. The sequence of the thin curves from top to bottom
corresponds to the set of values  $\xi$ = 0, 0.2, 0.4, 0.6, 0.8
and 1.0. These calculations are fulfilled for the Yukawa-potential
two-body model with the parameters (\ref{ndYukawaParameters}) of
the $nd$ system using the approximation (\ref{efrad}) when
$C_4=0$. The thick dashed curves are obtained from from an
analysis of the three-body results \cite{Ki95, Ki96} for the
Argonne $NN$ potential (AV18) taking the three-particle Urbana
forces (UR-IX) into account. The top dashed curve is for the $nd$
scattering, the bottom dashed curve is for the $pd$ scattering.
The lowest solid thick curve shows the calculations for the Yukawa
potential with the parameters (\ref{YukawaParameters}) for the
$pd$ system.} \label{fig4}
\end{figure}

In Fig. \ref{fig4} we depict the dependencies
$K_0^{(\xi)}(E_{\rm{lab}})$ on the energy in the range
$E_{\rm{lab}}\leq0.3$ MeV, where the Coulomb interaction effects
are  most pronounced. The calculations were performed with  the
set $\xi$ = 0, 0.2, 0.4, 0.6, 0.8 and 1.0; these values correspond
to the downward sequence of thin curves in Fig.4 (the top and
bottom curves are for $nd$ and $pd$ scattering, respectively). The
points on the curve correspond to the phase shift values found by
solving the Schr\"odinger equation in a continuum with the Coulomb
potential $\xi V_{\rm{C}}(r)$. The curves connecting the points
are the results of fitting using formula (\ref{efradE}). We took
the set of fitting parameters $ c_2=aC_2, c_4=aC_4$ and $E_0/a$,
which depends weakly on the scattering length $a$. It was found
that the fitted $aC_4$ values are unstable: they change abruptly
in magnitude with a variation in $\xi$ taking large values due to
the smallness of $k^4$ in the range of $E$ under consideration,
and even change their sign. This behavior is in agreement with the
conclusion of \cite{OrlOre06} according to which the parameter
$C_4$ cannot reliably be found without fitting the binding energy
of the $Nd$ system. In the range $E_{\rm{lab}}\leq0.3$ MeV, one
can assume that $aC_4$=0. Such an assumption was made in
\cite{Ki97} in determining  the $K_0(E)$ parameters for the $pd$
system. To fit the other parameters $a, aC_2$ and $E_0/a$ it is
sufficient to use $K_0^{(\xi)}(E)$ at low energies in the region
$E_{\rm{lab}}\leq 0.1$ MeV where the convergence of the expansion
is secured. It can be seen that the found parameters  for the
considered set of $\xi(E)$ make it possible to reproduce nicely
the $K_0^{(\xi)}(E)$ values calculated from the Schr\"odinger
equation at higher energies in the range under study.

It is remarkable that the observed effect of consideration of the
Coulomb interaction within the two-body model is in qualitative
agreement with the results obtained within  the three-body problem
for the  $nd$ and $pd$ systems. The corresponding dash-line curves
are shown in Fig. \ref{fig4} together with our results for
$K_0^{(\xi)}(E)$. There is very good agreement for the $nd$
scattering (the top curve). The significant observable differences
for the $pd$ scattering are related to the difference of the
close-to-zero value $^2a_{pd}$ = 0.024 fm in \cite{Ki97} from
0.136 fm, the value obtained in this study within the two-body
Yukawa-potential model.

The first estimate of the doublet $pd$ scattering length within
the two-body model was made in \cite{Tom87}; the result ($^2a_{pd}
\cong 0.15$ fm) is close to that obtained by us using the Yukawa
potential. Let us remember that the potentials considered in
\cite{Tom87} include three parameters fitted from the doublet $nd$
scattering length ($^2a_{nd} \cong 0.65$ fm) and the $^3$H and
$^3$He binding energies.

Although there is no reliable experimental result for $^2a_{pd}$
at present, we think that the three-body theoretical result in
\cite{Ki97} is currently the best estimate.

Therefore, for the $pd$ system one can fit the parameters of the
Yukawa and Hulth\'{e}n potentials by analogy with the fitting for
the $nd$ system: from the experimental binding energy
($\epsilon_\tau=7.73$ MeV) and the theoretical $pd$ scattering
length ($^2a_{pd}$ = 0.024 fm).

As a result of such fitting for the Yukawa potential we found the
following parameter values and the corresponding physical
quantities calculated by solving the Schr\"odinger equation for
the bound state and continuum (lengths in fm, energies in MeV)
\begin{equation} \label{YukawaParameters}
{(\rm{Yu},{pd})} \quad R=2.82,\quad  V_0=16.93; \quad {}^2a_{pd} =
0.026,\quad \epsilon_\tau=7.73, \quad C_{\tau}^2=3.30,\quad
\tilde{G}^2_\tau=1.30.
\end{equation}
Obviously one can ignore the difference in $^2a_{pd}$ (0.024 and
0.026). The values of ${C_\tau}^2$ and especially of the ${\tilde
G_\tau}^2$ in (\ref{YukawaParameters}) differ only  a little from
those reported in (\ref{ndYukawaParameters}) and in Table 1 for
the $nd$ system. We also calculated the features of the
subthreshold $pd$ resonance for the parameter set
(\ref{YukawaParameters}) using the effective-range function values
from the Schr\"odinger equation by this function fitting for the
set of the energy $E_{\rm{lab}}$ in the range (0.02--0.16) MeV
using the formula (\ref{efradE}). With the  parameters obtained in
this way and when $b_2=0$ $(b_1=1.216$ MeV$^{-1}$, $d_1=5.112$
fm/MeV) we found  $E_{\rm{res}}^{pd}=$ $=-(0.310+i0.074)$ MeV and
$\tilde{G}^2_{\rm{res}}=(-0.00463\pm i0.00215)$ fm which is in
reasonable agreement with of our analysis of the three-body
calculations \cite{Ki97} made by the Pisa group of  physicists:
$E_{\rm{res}}^{pd}=-(0.319+i0.099)$ MeV,
$\tilde{G}^2_{\rm{res}}=(-0.00596\pm i0.00358)$ fm.

The differences between the corresponding parameter pairs which
resulted from the breakdown of the charge independence are also
small, although they lead to a rather high change of the
scattering length, which is known as the value most sensitive to
parameter change.

We calculate the function $K_0^{(\xi)}(E)$ with the parameters
(\ref{YukawaParameters}) at $\xi=1$ (see Fig. \ref{fig4}). A
comparison with the three-body curve in the same figure shows that
in the energy region under study the two-body model with the
Yukawa potential when the parameters are fitted from the
three-body binding energy value and the doublet $pd$ scattering
length given in (\ref{YukawaParameters}) leads to a larger
difference between the values of the function $K_0^{(\xi)}(E)$ at
$\xi=0$ and $\xi=1$. This means that the role of the Coulomb
repulsion is somewhat overestimated. But the observable difference
between the two-body and three-body curves at $\xi=1$ is well
known to be smaller than the experimental error. In the low-energy
region considered  reliable measurements of the $pd$ scattering
actually does not exist due to the high Coulomb barrier.

With the parameters of the effective-range function
$K_0^{(\xi)}(E)$  found in such a way for the Yukawa potential for
the $nd$ system we calculate $k_{\rm{res}}(\xi)$ and
$\tilde{G}^2(\xi)$.

The charge dependence of the scattering length $^2a_{Nd}(\xi)$ (
the most important for the trajectory constructing) which we
calculate within the two-body Yukawa-potential model with the
parameters (\ref{ndYukawaParameters}) is shown in Fig. \ref{fig1}.
This function $^2a_{Nd}(\xi)$ is well approximated by a quadratic
polynomial
\begin{equation} \label{aNd-xi}
^2a_{Nd}(\xi)=0.667-0.273\xi-0.244\xi^2.
\end{equation}

The changing of the scattering length with the charge variation is
quite sudden: the ratio of the scattering lengths at $\xi=0$ and
$\xi=1$ is close to 4, whereas the corresponding ratio in the case
of the $Np$ scattering is near 3. The changing of the reciprocal
quantity $[^2a_{Nd}(\xi)]^{-1}$ is quicker. It can be described by
the exponentially increasing function at $\xi\rightarrow1$ or as a
function with a pole at $\xi>1$. It should be  stressed here that
the relation (\ref{Brown}) is by no means universal. Obviously,
this is not valid for the system in which the effective-range
function has a pole close to the scattering channel threshold.
However, because of the scale difference, the changing  of the
$Nd$ scattering length with the charge variation looks much weaker
in Fig. \ref{fig1} compared with the dependence on $\xi$ of the
$Np$ scattering length. The corresponding curves differ not only
quantitatively but also qualitatively.

The corresponding trajectories of the subthreshold resonance
momentum and of the renormalized vertex constant for a transition
from the spin-doublet proton-deuteron system to the
neutron-deuteron system are shown in Figs. \ref{fig2} and
\ref{fig3}. Explicit qualitative and quantitative differences are
seen between the corresponding trajectories for the
nucleon-nucleon and nucleon-deuteron systems. In  the latter case,
at a transition from the resonance to the virtual state, the pole
trajectory is shifted not upward but downward and is more remote
from the physical energy range. The $\mid\tilde{G}^2\mid$
magnitude for the $Nd$ system is about factor 5 smaller than for
the $Np$ system (for clearness in Fig. \ref{fig3} the trajectory
(5 $\tilde{G}^2$) is also shown for $Nd$ system). The charge
dependence of $\tilde{G}^2$ also differs from the case of
nucleon-nucleon system.

Due to the correlation between $E_0$ and $^2a_{pd}$, the
three-body $E_0$ value also significantly differs from the
two-body model result. There is a large difference between the
values of parameter $E_0/^2a_{pd}$ which is less sensitive to a
change in the scattering length $^2a_{pd}$. The $E_0/^2a_{pd}$ =
0.13  MeV/fm was obtained in the three-body approach \cite{Ki97}
whereas the value of  $E_0/^2a_{pd}\approx0.2$ MeV/fm was found in
the present study on the basis of the two-body model with the
Yukawa potential.

\section{\boldmath THE ${\mathit p}$-WAVE RESONANCES IN THE SCATTERING OF THE NUCLEON BY
$^4\rm{He}$}

The resonances in the $N\alpha$ scattering near the threshold have
been studied in many experimental and theoretical works (see, for
example, [36] and references therein). In [36] the results of the
R-matrix analyses were given not only for the elastic scattering
but also for the stripping and pick-up reactions. The
corresponding results for the resonant energies and widths are
also presented together with the experimental errors. It should be
noted that the R-matrix approach used contains a considerable
number of parameters. Including data on nuclear reactions  in the
analysis leads to additional errors due to its dependence on the
model used to describe the reaction considered.

As we note in the introduction, it seems that the effective-range
expansion was used for the first time in \cite{Ahmed76} to find
the pole positions on the unphysical energy sheet which correspond
to the resonances in the $p$-wave amplitude of the $N\alpha$
elastic scattering. We limit ourselves in the present work to
considering these $p$-wave resonances when formula
(\ref{CoulombKl}) has the following form:
\begin{equation} \label{CoulombK1}
K_1(k^2) = k^3 (1+\gamma^2)\left[[2\pi\gamma/[\exp(2\pi\gamma)-1]]
\cot\delta_l^C(E) +\gamma[\Psi (1+i\gamma )+\Psi (1-i\gamma
)-2\ln\gamma]\right].
\end{equation}
For the renormalized VC we correspondingly derive the expression:
\begin{equation} \label{Gren2dif1}
{\tilde{G}_1}^2 =
\frac{(-2\pi/\mu^2)p^3}{\frac{d}{dk}\left[K_1(k^2)-(Q(k)+i{C_1}^2k^3)\right]_{k=p}}
,
\end{equation}
where the notations are the same as in \cite{Ahmed76}:
\begin{equation}\label{CoulombQ1}
Q(k)=\gamma (1+\gamma^2) k^3
[\Psi(i\gamma)+\Psi(-i\gamma)-2\ln\gamma],
\end{equation}
\begin{equation}\label{CoulombC1}
C_1^2=2\pi\gamma (1+\gamma^2) /[\exp(2\pi\gamma)-1].
\end{equation}

The results of our calculations are compared in Table \ref{table2}
with those obtained in \cite{Ahmed76} and \cite{Safron83} (see
also \cite{Blokh84}) as well as in the more recent work
\cite{Efros96} where the experimental data on the radiation
capture of the deuteron by $^3$H and $^3$He was analyzed. As far
as we know, the renormalized VCs for the resonances discussed here
were previously calculated only in \cite{Safron83}  using the
$N/D$ method and also presented in \cite{Blokh84}. In Table
\ref{table2} the quantity $R_1$ for the residue of the T-matrix at
the pole is given as well. It is connected with the renormalized
${\tilde{G}_l}^2$ by the relationship (see \cite{Blokh84})
\begin{equation} \label{SafronRl}
{\tilde{G}_l}^2=(-1)^l 2 \pi \kappa \mu^{-2} R_l.
\end{equation}

\begin{table}
\caption{The complex values of the resonance CM energies $E_r
-i\Gamma/2$ (in  MeV) for $j^p=3/2^-$ and $j^p=1/2^-$ relative to
the threshold in the channel $\alpha+N$ ($^5$He, $^5$Li ) and of
the $T$ matrix residues $R_1=|R_1|\rm{exp}(\it i\varphi_1)$,
($\varphi_1$ in deg.).\label{table2}}.
\begin{ruledtabular}
\begin{tabular}{|lcccccc|}
\,\,Method &Nucleus&$j^p$&$E_r$&$\Gamma$&$|R_1|$&$\varphi_1$\,\,\\
\hline
\,\,Eff. rad. \cite{Ahmed76}&& &0.778&0.639&-&-\,\,\\
\,\,$N/D$ \cite{Safron83}, \cite{Blokh84}&& &0.697&0.542&0.160&132\,\,\\
\,\,$(d,\gamma)$ \cite{Efros96}&$^5$He&$3/2^-$&0.80&0.65&-&-\,\,\\
\,\,$R$-function  \cite{Bond77}&&&0.771&0.644&-&-\,\,\\
\,\,$R$-matrix  \cite{Csoto97}&&&0.80&0.65&-&-\,\,\\
\,\,Eff. rad. (present paper); &&&0.778&0.641&0.171&-133\,\,\\
\hline
\,\,Eff. rad.   \cite{Ahmed76}&& &1.999&4.534&-&-\,\,\\
\,\,$N/D$ \cite{Safron83}, \cite{Blokh84}&& &1.875&5.296&0.237&177\,\,\\
\,\,$R$-function \cite{Bond77}&$^5$He&1/2$^-$&1.970&5.218&-&-\,\,\\
\,\,$R$-matrix \cite{Csoto97}&&&2.07&5.57&-&-\,\,\\
\,\,Eff. rad. (present paper) &&&2.000&4.533&0.199&178\,\,\\
 \hline
\,\,Eff. rad. \cite{Ahmed76}&&&1.637&1.292&-&-\\
\,\,$N/D$ \cite{Safron83}, \cite{Blokh84}&&&1.655&1.278&0.304&-136\,\,\\
\,\,$(d,\gamma)$ \cite{Efros96}&$^5$Li&$3/2^-$&1.72&1.28&-&-\,\,\\
\,\,$R$-matrix \cite{Csoto97}&&&1.69&1.23&-&-\,\,\\
\,\,Eff. rad. (present paper); &&&1.630&1.437&0.278&-148\,\,\\
\hline
\,\,Eff. rad.  \cite{Ahmed76}&& &2.858&6.082&-&-\,\,\\
\,\,$N/D$   \cite{Safron83}, \cite{Blokh84}&$^5$Li&1/2$^-$
&2.691&6.448&0.313&-178\,\,\\
\,\,$R$-matrix \cite{Csoto97}&&&3.18&6.60&-&-\,\,\\
\,\,Eff. rad. (present paper); &&&2.34&6.01&0.287&4.2\,\,\\
\end{tabular}
\end{ruledtabular}
\end{table}

A comparison of our results with those presented in \cite{Blokh84}
for the $N/D$ method shows some differences in the resonance
positions and their residues. The absolute values of the residues
more or less agree with each other, while the angles $\varphi$
almost coincide in two cases but  differ in the other two. The
$\varphi$ values for the states $^5$He ($j=1/2$) and
$^5$Li($j=3/2$) are in good agreement whereas for the state $^5$He
($j=3/2$) the angles $\varphi$ differ mainly by their sign. This
means that the residues are a complex conjugate. But for the state
$^5$Li ($j=1/2$) the difference between them amounts to about 180
deg. This means that the residues  differ approximately in their
signs.

Taking the Coulomb interaction into account leads to the sign
change of the angles given in \cite{Blokh84}. In our calculations,
the corresponding  change of $\varphi$  more correlates with the
value of the total angular momentum $j$  whereas  the magnitude of
the residue  does not depend strongly on $j$. The inclusion of the
Coulomb interaction leads to a marked increase in the absolute
value $|R_1|$ of the residue. Apparently, the noted differences
between the $\varphi$ values are not fundamental. It is known that
resonance poles exist in pairs which are mirror symmetrical
relative to the imaginary axis of the momentum, the corresponding
residues being a complex conjugate. The sign difference of the
residues also does not change the input of the poles into the
cross section if there is no essential interference with a
nonresonant amplitude.

\begin{figure}[bp]
\resizebox*{0.7\textwidth}{!}{\includegraphics{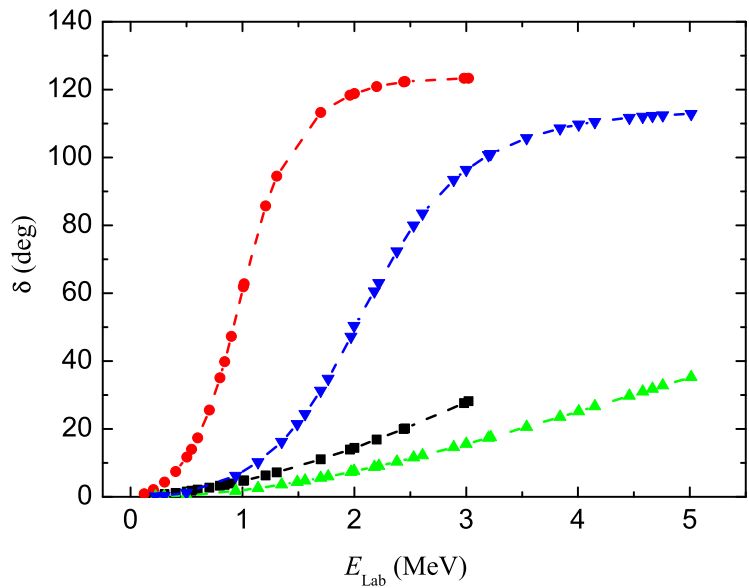}}
\caption{The results of the phase shift analysis \cite{Arndt73B}
of the nucleon scattering by the nucleus $^4$He for the $p$-wave:
the solid circles and squares are for $n\alpha$ scattering when
$j=3/2$ and $j=1/2$ respectively, while the inverted and right
triangles  stand for $p\alpha$ scattering when $j=3/2$ and
$j=1/2$, respectively.} \label{fig5}
\end{figure}

In \cite{Ahmed76} the phase shift analysis  of the $N\alpha$
scattering presented in \cite{Arndt73A} was used. We compare two
variants of the phase shift analysis of the $N\alpha$ scattering
which  differ in their maximal energy values: (A) $E_{\rm{cm}}
\leq3$ MeV for the $n\alpha$ scattering and $E_{\rm{cm}} \leq$ 5
MeV for the $p\alpha$  scattering \cite{Arndt73B} and (B)
$E_{\rm{lab}}\leq21$ MeV \cite{Arndt73A}. We include in Table
\ref{table2} only our results for variant B. For the states when
$j=1/2$ the area of the fast decrease of the phase shift is
situated at a higher energy (see Fig. \ref{fig5}). Furthermore,
the resonance width is almost ten times greater than that for the
resonance when $j=3/2$. At the same time, the phase shift does not
pass  the value $\pi/2$.

The short energy interval in  variant A does not cover the area
where  the phase shift  decreases quickly due to the resonance
pole position. Due to this, variant A does not permit a correct
reproduction of the parameters for the effective-range function.
When $j=3/2$ the short interval is representative enough to give
reasonable parameters for $K(k^2)$ and for the corresponding
resonances. The increase of the energy interval for the fitting
parameters  up to  energy 21 MeV leads to slight differences in
the results when $j=3/2$. This testifies to the fact that the
resonance poles considered are situated in the area where the
effective-range expansion converges.

A comparison of our results with those in \cite{Ahmed76} shows
excellent agreement for the $n\alpha$ scattering but considerable
disagreement for the $p\alpha$ scattering. We believe that our
calculations are more accurate due to the progress in calculation
methods  for equation solving, including the function $\psi$. In
Table \ref{table2} the complex resonance energies calculated using
the R-matrix method are generally in reasonable agreement with
those obtained by other methods (see also \cite{Csoto97}).

It should be noted that for the $p\alpha$ scattering the agreement
of our results using the analytical continuation of the
effective-range function with other results published in the
literature is not as good as that between the latter and the
results in \cite{Ahmed76}.

\section{CONCLUSIONS}

Thus, in this study the correct expressions are derived for the
renormalized vertex constant for the decay of a nucleus into two
charged particles in a state with the arbitrary angular orbital
momentum within the effective-range theory. However, it should be
noted that, since this theory is valid only for low enough
energies when an expansion of the effective-range function in
powers of $k^{2}$ converges, there is also a limitation on the
upper value of the orbital angular momentum. It is likely we can
largely limit ourselves to considering   $s$ and $p$-waves. For
the $s$-wave ( see also \cite{Erem07}) we consider both the
standard effective-range expansion and the expansion with a pole
of the effective-range function, which is necessary in the case of
the $Nd$ system. The formulas derived are applied to the bound
state of $^{3}$He nucleus and to the subthreshold resonance in the
$pd$ scattering which corresponds to the virtual level of $^{3}$H
and also to the resonance of the  $pp$ system. Trajectories are
constructed in the complex planes of the momentum and of the
renormalized vertex constant (real for neutral particles) at the
transitions from the subthreshold resonance in the ground singlet
$s$ state of the $pp$ system to the antibound $np$ system and from
the excited doublet state of the $pd$ system to the virtual triton
state at a gradual decrease in the Coulomb interaction to zero.
The constructed trajectories demonstrate the general physical
nature of the corresponding states, which differ only in the
Coulomb interaction.

The significant differences between the corresponding trajectories
for the $Np$ and $Nd$ system are related to both the larger
remoteness of the virtual triton pole and to the different
behavior of the scattering lengths $a(\xi)$ that depend more
strongly on charge: with $\xi$ increasing $a(\xi)(<0)$ quickly
increases for $Np$ scattering whereas $a_{Nd}(\xi)(>0)$ even more
quickly decreases for $Nd$ scattering. Because of the different
scales of variation, the change for the $Nd$ system looks slow and
almost a linear decrease in Fig. \ref{fig1}. It is remarkable that
the relatively sharp decrease in $a_{Nd}(\xi)$ with $\xi$
increasing due  only to the Coulomb interaction, which leads to a
reasonable value close to zero, was obtained by us within the
simple two-body model with two parameters, fitted from the triton
binding energy and doublet $nd$ scattering length. The doublet
$pd$ scattering length found is in good agreement with the results
of modern three-body calculations. The first result along these
lines was obtained in \cite{Tom87} where the three-parameter
two-body model was proposed (the third parameter was fitted from
the binding energy of $^3$He). The success of the two-body model
in a qualitative reproduction of the doublet scattering length
means that  the internal deuteron structure is apparently of no
importance at the nucleon and deuteron zero relative energy.

The parameter fitting  for the Yukawa potential from the
three-body $pd$ scattering length $^2a_{pd}$ considerably improves
the agreement between the two-body and three-body effective-range
functions at very low energies  $E_{\rm{lab}} \leq 0.2 $ MeV.

The reason for the difference in the trajectories for the $Np$ and
$Nd$ systems is that a singlet deuteron is in the ground state
with zero spin, for which the $np$ interaction differs from that
in the spin-triplet state, leading to the bound deuteron. The
ground state of an $Np$ system can be described within the
standard effective-range approximation. The virtual triton
corresponds to the excited state in which all quantum numbers,
except for the principal quantum number, are the same for the real
and virtual tritons. So for the virtual state one has the same
effective nuclear potential as for the bound state, and the
function $K_0^{(\xi)}(k^2)$ has a pole near the threshold.

The $\mid\tilde{G}^2\mid$ value for  $Nd$ system is smaller than
for $Np$ by a factor of about 5. Along with the larger remoteness
of the pole position from the physical region, this fact explains
why the $Nd$ subthreshold poles are much more difficult to find in
experiments in comparison with the well-studied $Np$ scattering
poles. The trajectory of $k_{\rm{res}}(\xi)$ which follows from
the approximate formula by Landau--Smorodinsky \cite{Brown} and
Schwinger \cite{Schwinger50}, relating $pp$ and $np$ scattering
lengths, is in good agreement with our result for the singlet $Np$
system in the model with the Yukawa potential. The difference is
due to using the nuclear potential for the $np$ system which leads
to a small change of the $Np$ scattering length at $\xi=1$ in
comparison with the experimental $pp$ scattering length.

Finally, we recalculated  the resonance positions
$E_{\rm{res}}-i\Gamma/2$ in the nucleon scattering on $^4$He in
the $p$-wave within the effective-range approximation and found
the corresponding residues at the poles which were then compared
with the results  [37,25] obtained in the \textit{N/D} method.

For the $p\alpha$ scattering when $j=1/2$ the $E_{\rm{res}}$ value
decreased slightly and when $j=3/2$  the $\Gamma$ value increased
a little in comparison with the results of the work \cite{Ahmed76}
where the effective-range approximation was also used.

The expressions obtained and the results of the numeral
calculations of the vertex constants for nucleus decay into two
charged fragments can be applied to the theory of reactions using
Feynman diagrams for a  description of the process mechanism, and
in the analysis of astrophysical nuclear synthesis reactions.

\section*{ACKNOWLEDGMENTS}
This study was partly supported by the Russian Foundation for
Basic Research, project no. 07-02--00609, and Grant NSh-485.2008.2
of the President of the Russian Federation for Support of Leading
Scientific Schools.


\begin{thebibliography}{99}
\bibitem{Schapiro}
I.\,S.~Shapiro, Zh. Eksp. Teor. Fiz. \textbf{41}, 1616 (1961)
[Sov. Phys. JETP \textbf{14}, 1148 (1962)]; \textit{Theory of
Direct Nuclear Reactions }(Atomizdat, Moscow, 1963) [in Russian];
Usp. Fiz. Nauk \textbf{92}, 549 (1967) [Sov. Phys. Usp.
\textbf{10}, 515 (1967)].

\bibitem{Erem07}
V.\,O.~Eremenko, L.\,I.~Nikitina, and Yu.\,V.~Orlov, Izv.  Akad.
Nauk, Ser. Fiz. \textbf{71}, 819 (2007)[Bull. Russ. Akad. Sci.:
Physics \textbf{71}, 791 (2007)].

\bibitem{Orl09}
Yu.\,V.~Orlov, V.\,O.~Eremenko, B.\,F.~Irgaziev, and
L.\,I.~Nikitina, Izv. Akad. Nauk, Ser. fiz.  \textbf{73}, 826
(2009). [Bull. Russ. Russ. Akad. Sci.: Physics \textbf{73}, 778
(2009)]

\bibitem{Yarmukh07}
S.\,B.~Igamov, R.~Yarmukhamedov, Nucl. Phys. A \textbf{781}, 247
(2007).

\bibitem{Mur83}
V.\,D.~Mur, A.\,E.~Kudryavtsev, and V.\,S.~Popov, Yad. Fiz.
\textbf{37}, 1417 (1983) [Sov. J. Nucl. Phys. \textbf{37}, 844
(1983)].

\bibitem{Delv60}
L.\,M.~Delves, Phys. Rev.  \textbf{118}, 1318 (1960).

\bibitem{Oers67}
W.\,T.\,H.~Van Oers and J.\,D.~Seagrave, Phys. Lett. B
\textbf{24}, 562 (1967).

\bibitem{Whit76}
J.\,S.~Whiting and  M.\,G.~Fuda, Phys. Rev. C \textbf{14}, 18
(1976).

\bibitem{Sim87}
I.\,V.~Simenog, A.\,I.~Sitnichenko, and D.\,V.~Shapoval, Yad. Fiz.
\textbf{45}, 60 (1987) [Sov. J. Nucl. Phys. \textbf{45}, 37
(1987)].

\bibitem{VanHaer}
H.\,van~Haeringen, \textit{Charged-Particle Interactions}
(\textit{Theory and Formulas})  (Coulomb Press, Leyden, 1985).

\bibitem{Kok80}
L.\,P.~Kok, Phys. Rev. Lett.  \textbf{45}, 427 (1980).

\bibitem{Arv74}
J. Arvieux, Nucl. Phys.  A \textbf{221}, 253 (1974).

\bibitem{Baz71}
A.\,I.~Baz', Ya.\,B.~Zel'dovich, and A.\,M.~Perelomov,
\textit{Scattering, Reactions and Decays in Nonrelativistic
Quantum Mechanics }(Nauka, Moscow, 1971) [in Russian].

\bibitem{Sit75}
A.\,G.~Sitenko \textit{Scattering Theory }(Vishcha Shkola, Kiev,
1975) [in Russian].

\bibitem{Landau63}
L.\,D.~Landau and E.\,M.~Lifshitz, \textit{Course of Theoretical
Physics,} Vol. 3: \textit{Quantum Mechanics: Non-Relativistic
Theory} (Nauka, Moscow, 1989; Pergamon, Oxford, 1977).

\bibitem{Ahmed76}
M.\,U.~Ahmed and P.\,E.~Shanley,  Phys. Rev. Lett. \textbf{36}, 25
(1976).

\bibitem{Brown}
G.\,E.~Brown and A.\,D.~Jackson,  \textit{Nucleon-Nucleon
Interactions} (North-Holland, Amsterdam, 1976; Atomizdat, Moscow,
1979).

\bibitem{OrlIzv05}
Yu.\,V.~Orlov,  Izv. Akad. Nauk, Ser. Fiz. \textbf{69}, 144 (2005)
[Bull. Russ. Akad. Sci.: Physics \textbf{69}, 149 (2005)].

\bibitem{OrlOre06}
Yu.\,V.~Orlov and  Yu.\,P.~Orevkov, Yad. Fiz. \textbf{69}, 855
(2006) [Phys. At. Nucl. \textbf{69}, 828 (2006)].

\bibitem{Ki95}
A.~Kievsky, M.~Viviani, and S.~Rosati, Phys. Rev.  C \textbf{52},
R15 (1995).

\bibitem{Ki96}
A.~Kievsky, S.~Rosati   \textit{et al.}, Nucl. Phys.  A
\textbf{607}, 402 (1996).

\bibitem{Csoto}
A.~Csoto and G.\,M.~Hale, Phys. Rev.  C \textbf{59}, 1207 (1999);
(Erratum) \textbf{62}, 049901 (2000).

\bibitem{Gold}
M.\,L.~Goldberger, K.\,M.~Watson, \textit{Collision Theory}
(Wiley, New York, 1967; Mir, Moscow, 1967).

\bibitem{Janke}
E.~Janke, F.~Emde, and F.~L\"osch, \textit{Tables of Higher
Functions }(McGrawHill, New York, 1960; Nauka, Moscow, 1964).

\bibitem{Blokh84}
L.\,D.~Blokhintsev, A.\,M.~Mukhamedzhanov, and A.\,N.~Safronov,
Fiz. Elem. Chastits At. Yadra \textbf{15}, 1296 (1984) [Sov. J.
Part. Nucl.\textbf{15}, 580 (1984)].

\bibitem{Blokh77}
L.\,D.~Blokhintsev, I.~Borbely, and  E.\,I.~Dolinskii, Fiz. Elem.
Chastits At. Yadra \textbf{8}, 1189 (1977)[Sov. J. Part. Nucl.
\textbf{8}, 485 (1977)].

\bibitem{Schwinger50}
J.~Schwinger, Phys. Rev. \textbf{78}, 135 (1950).

\bibitem{OrlOreNik02}
Yu.\,V.~Orlov,  Yu.\,P.~Orevkov, and  L.\,I.~Nikitina, Yad. Fiz.
\textbf{65}, 396 (2002) [Phys. At. Nucl. \textbf{65}, 371 (2002)].

\bibitem{Friar82}
J.\,L.~Friar, B.\,F.~Gibson, D.\,R.~Lehman, and G.\,L.~Payne,
Phys. Rev. C \textbf{25}, 1616 (1982).

\bibitem{Ki97}
A.~Kievsky, S.~Rosati, M.~Viviani  \textit{et al.}, Phys. Lett.  B
\textbf{406}, 292 (1997).

\bibitem{Saf89}
A.\,N.~Safronov, Yad. Fiz. \textbf{50}, 951 (1989) [Sov. J. Nucl.
Phys. \textbf{50}, 593 (1989)].

\bibitem{Petr88}
N.\,M.~Petrov, Yad. Fiz. \textbf{48}, 50 (1988) [Sov. J. Nucl.
Phys. \textbf{48}, 31 (1988)].

\bibitem{Tom87}
L.~Tomio, A.~Delfino, S.\,K.~Adhikari, Phys. Rev.  C \textbf{35},
441 (1987).
\bibitem{Irg06}
B.\,F.~Irgaziev,  L.\,I.~Nikitina, and Yu.\,V.~Orlov, Izv. Akad.
Nauk, Ser. Fiz. \textbf{70}, 227 (2006) [Bull. Russ. Akad. Sci.:
Physics \textbf{70}, 257 (2006)].

\bibitem{Orl06}
Yu.\,V.~Orlov and  L.\,I.~Nikitina, Yad. Fiz. \textbf{69}, 631
(2006) [Phys. At. Nucl. \textbf{69}, 607 (2006)].

\bibitem{Csoto97}
A.~Csoto and G.\,M.~Hale, Phys. Rev.  C \textbf{55}, 536 (1997).

\bibitem{Safron83}
A.\,N.~Safronov, Pis'ma Zh. Eksp. Teor. Fiz. \textbf{37}, 608
(1983) [JETP Lett. \textbf{37}, 727 (1983)].

\bibitem{Efros96}
V.\,D.~Efros and H.~Oberhummer, Phys. Rev.  C \textbf{54}, 1485
(1996).

\bibitem{Bond77}
J.\,E.~Bond and F.\,W.\,K.~Firk, Nucl. Phys.  A \textbf{287}, 317
(1977).

\bibitem{Arndt73A}
R.\,A.~Arndt and L.\,D.~Roper, Nucl. Phys.  A \textbf{209}, 447
(1973).

\bibitem{Arndt73B}
R.\,A.~Arndt and L.\,D.~Roper, Nucl. Phys.  A \textbf{209}, 429
(1973).

\end{thebibliography}
\end{document}